\documentstyle[prl,aps,multicol,psfig]{revtex}
\begin{document}

\draft \wideabs{
\title{Stripe Conductivity in  La$_{1.775}$Sr$_{0.225}$NiO$_{4}$}
\author{Yu.G.Pashkevich,$^1$ V.A.Blinkin,$^2$ V.P.Gnezdilov,$^3$
V.V.Tsapenko,$^3$ V.V.Eremenko,$^3$ P.Lemmens,$^4$ M.Fischer,$^4$
M.Grove,$^4$ G.G\"{u}ntherodt,$^4$ L.Degiorgi,$^5$ P.Wachter,$^5$
J.M.Tranquada,$^6$ and D.J.Buttrey$^7$}
\address{$^1$A.A.Galkin Donetsk Phystech NASU, 83114 Donetsk, Ukraine \\
$^2$Inst. for Single Crystals NASU, 310001 Kharkov, Ukraine\\
$^3$B.I. Verkin Inst. for Low Temp. Physics NASU, 310164 Kharkov,
Ukraine\\ 
$^4$2. Physikalisches Institut, RWTH Aachen, 52056 Aachen, Germany\\ 
$^5$Laboratorium f\"{u}r Festk\"{o}rperphysik,
ETH-Z\"{u}rich, CH-8093, Z\"{u}rich, Switzerland\\ 
$^6$Brookhaven National Laboratory, Upton, NY 11973\\ 
$^7$University of Delaware, Newark, Delaware 19716}
\date{\today}
\maketitle

\begin{abstract}
We report Raman light-scattering and optical conductivity
measurements on a single crystal of
La$_{1.775}$Sr$_{0.225}$NiO$_{4}$ which exhibits incommensurate
charge-stripe order.  The extra phonon peaks induced by stripe
order can be understood in terms of the energies of phonons that
occur at the charge-order wave vector, ${\bf Q_c}$.  A strong Fano
antiresonance for a Ni-O bond-stretching mode provides clear
evidence for finite dynamical conductivity within the charge
stripes.
\end{abstract}
 }

\narrowtext

Recent experiments have shown a difference in the conductivity behavior of
the stripe ordered phases in the cuprates 
La$_{2-x-y}$Nd$_{x}$Sr$_{y}$CuO$_{4}$ and the nickelates
La$_{2-x}$Sr$_{x}$NiO$_{4}$. Strong localization and binding of charges to
the lattice in nickelates manifest themselves by the appearance of: 1) an
insulating state and a charge gap in optical conductivity
\cite{ido91,kats96}; and 2) additional diffraction peaks due to ionic
displacements that are induced below $T_{c},$ the charge ordering
temperature \cite{chen93,tran94,lee97}. In cuprates such a gap is not
observed \cite{taj98} even though charge-order superlattice peaks have been
detected by both neutron and x-ray scattering \cite{1jt95}. The diffraction
measurements have shown that the corresponding lattice modulation is much
smaller in cuprates than in nickelates \cite{zim98,tran96}. At the same time
both cuprates and nickelates have demonstrated incommensurate magnetic
ordering \cite{1jt95,1jt96,vigl97}. This by itself is important evidence of
the stripe state because it implies a modulation of the charge density.
Moreover, the coexistence of superconductivity and stripe order has been
observed in La$_{1.6-x}$Nd$_{0.4}$Sr$_{x}$CuO$_{4}$ with $x=0.12$, 0.15, and
$0.20$ \cite{2jt97,ost97}.

The problem of whether or not stripes in cuprates and nickelates are
insulating or metallic is fundamental to the physics of the stripe state.
One can expect two possible scenarios which could lead to conductivity in
the stripe state. According to the first one the stripes themselves are
insulating but the system can be metallic due to fluctuations and motion of
stripes \cite{1zaa99}. Alternatively, metallic conductivity may exist along
the charge threads without a violation of stripe ordering as a whole. In the
latter case, Coulomb interactions between neighboring stripe should lead to
charge-density-wave order along the stripes at sufficiently low temperatures
and in the absence of stripe fluctuations \cite{kive98}.

Previous studies have shown that the stripe-order in La$_{2-x}$Sr$_x$NiO$_4$
with $x=\frac13$ has a short-period commensurability \cite{lee97} and a very
large charge gap of 0.26~eV relative to the charge ordering temperature of
230~K \cite{kats96}. Thermodynamic measurements have indicated that 
$x=\frac13$ may be somewhat special \cite{rami96}, while recent phonon
density-of-states measurements have found doping-dependent changes of the
in-plane Ni-O bond-stretching modes for $x=\frac14$, $\frac13$, and $\frac12$
\cite{mcqu99}. Here we present results from Raman light scattering (RLS) and
optical conductivity measurements on a single crystal with $x=0.225$, which
was previously characterized by neutron \cite{1jt96} and x-ray \cite{vigl97}
diffraction. The stripe order in this sample is incommensurate, and the hole
density per Ni site along a stripe is significantly less than 1 (in contrast
to $x=\frac13$, where the density is exactly 1). We observe a strong Fano
antiresonance in the optical conductivity at an energy well below the charge
pseudo-gap, which is 0.105 eV at 10~K. From a careful analysis of the phonon
spectra, we conclude that the energy of the antiresonance corresponds to
Ni-O bond stretching motions along the stripes. It follows that the
antiresonance, which results from electron-phonon coupling, provides strong
evidence for finite conductivity along the stripes, at least at
optical-phonon frequencies.

The light scattering measurements were carried out in quasi-backscattering
geometry using 514.5-nm argon laser light. The incident laser beam of 10 mW
or 15 mW power was focused onto a 0.1 mm diameter spot on the $a$-$b$ plane
of the mirror-like polished crystal surface. The $x,y,z$- crystallographic
axes in the $I4/mmm$ setting were determined by x-ray Laue diffraction. The
incident photons were polarized along or perpendicular to the 
$x^{\prime}=x+y,$ or $y^{\prime}=-x+y$ diagonal directions between in-plane
Ni-O bonds. The scattered photons were polarized either parallel or
perpendicular to the incident photons. Infrared (IR) reflectivity
measurements in the 25--$10^{5}$~cm$^{-1}$ frequency region were carried out
in the geometry of normal incidence to the $a$-$b$ plane of the sample.
Optical conductivity spectra, $\sigma(\omega)$, were obtained by
Kramers-Kronig analysis of the reflectivity data.

Let us begin with the optical conductivity spectra shown in Fig.~1. The
conductivity below 2000 cm$^{-1}$ clearly decreases as the temperature is
lowered. If we linearly extrapolate the frequency dependence from above 1500
cm$^{-1}$, as indicated by the dashed line for the 10~K spectrum, then,
following Katsufuji {\it et al.} \cite{kats96}, we estimate a
low-temperature gap of 840 cm$^{-1}$. At 150~K, which is the charge ordering
temperature ($T_c$) indicated by diffraction \cite{1jt96,vigl97}, the
conductivity has changed little. The extrapolated gap reaches zero somewhat
closer to 200~K. Note that at low temperature, although $\sigma(0)\approx0$,
the residual conductivity within the extrapolated gap is significantly
higher than that found for $x=\frac13$ \cite{kats96}.

\begin{figure}
\centerline{\psfig{figure=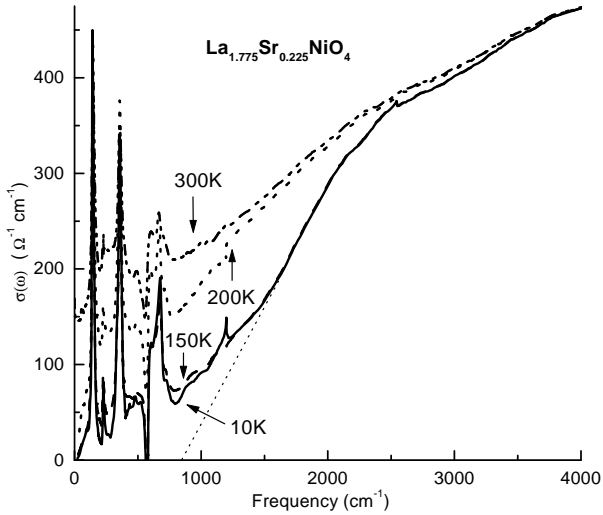,width=8cm,bbllx=14pt,bblly=365pt,%
bburx=555pt,bbury=826pt}} 
\caption{The temperature dependence of the optical conductivity spectra of
La$_{1.775}$Sr$_{0.225}$NiO$_4$ below 4000~cm$^{-1}$ for incidence
normal to the $a$-$b$ plane.}
\end{figure}

The temperature-dependent Raman spectra for two polarizations are presented
in Fig.~2. For both $x^{\prime }x^{\prime }$ and $x^{\prime }y^{\prime }$
polarizations there is a significant electronic background that shifts from
low to high energy as the temperature decreases. Low-temperature scans in the
$x^{\prime }y^{\prime }$ geometry reveal 2-magnon scattering bands at
739~cm$^{-1}$ and 1130~cm$^{-1}$, frequencies remarkably similar to those in
La$_{5/3}$Sr$_{1/3}$NiO$_4$ \cite{blum,yama98}. For both $x=0.225$ and
$x=\frac13$, the 2-magnon features disappear at approximately the
charge-ordering temperature determined by neutron diffraction, 150~K and
240~K, respectively \cite{lee97,1jt96}. Analysis of the 2-magnon spectra will
be presented elsewhere.

Next we consider the phononic features. In Fig.~2, the numberic
labels denote the positions of lines (in cm$^{-1}$) at low
temperature, determined by curve fitting. An expanded view of the
low-frequency range of the optical conductivity is shown in
Fig.~3; again, the numbers label low-temperature peak positions.

\begin{figure}
\centerline{\psfig{figure=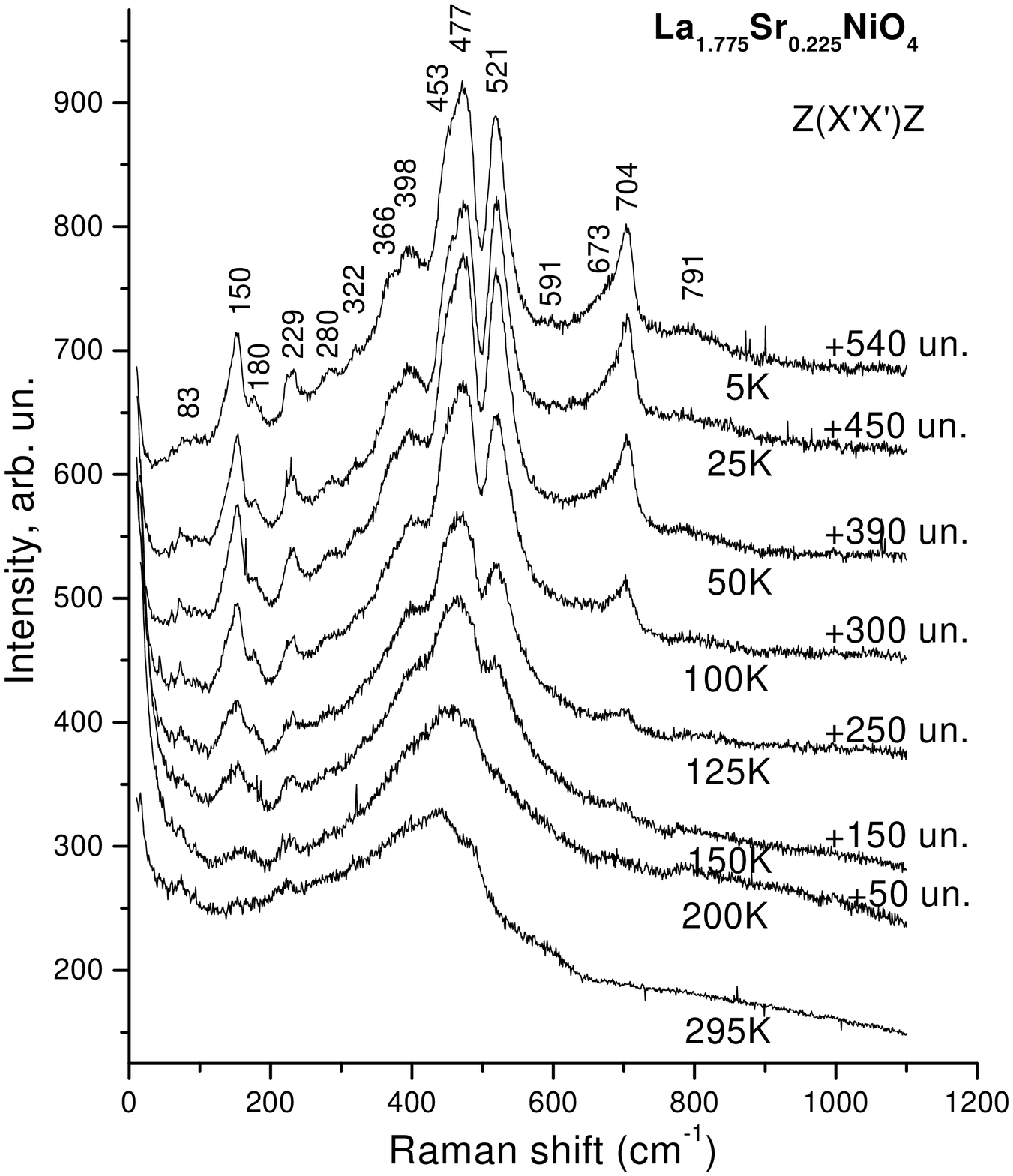,width=8cm,bbllx=14pt,bblly=175pt,%
bburx=570pt,bbury=825pt}}
\centerline{\psfig{figure=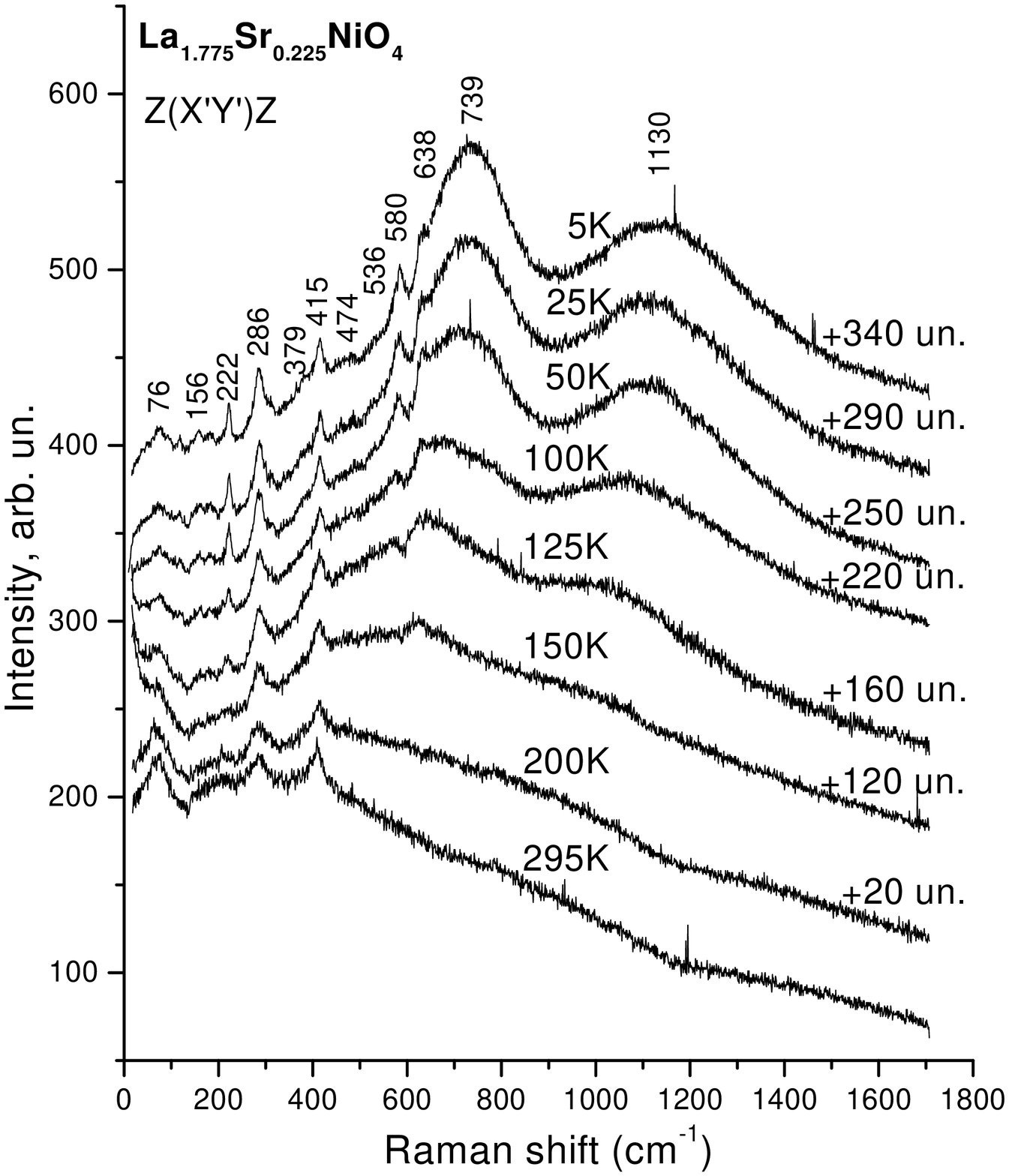,width=8cm,bbllx=14pt,bblly=175pt,%
bburx=570pt,bbury=825pt}} 
\caption{Temperature dependent Raman
light scattering spectra for $x'x'$ and $x'y'$ polarization of
single crystal La$_{1.775}$Sr$_{0.225}$NiO$_4$. The vertical offsets
are noted on the right. }
\end{figure}

To interpret the spectra, we begin by noting that RLS and IR measurements
are sensitive only to modes with momentum transfer $\hbar Q=0$. One
certainly expects a significant contribution from optically-active phonon
modes corresponding to the average lattice structure. In addition, the
occurrence of stripe order, with a characteristic wave vector ${\bf Q}_c$,
lowers the translational symmetry, and must lead to the appearance of extra
lines in the RLS and IR spectra. Finally, the Sr dopant ions locally break
the lattice symmetry, and hence can induce extra features, which are
expected to be temperature independent.

The symmetry of the mean tetragonal lattice is described by space group 
$I4/mmm$. The corresponding Raman-active phonons are distributed among the
irreducible representations (IREPs) of the space group as $2A_{1g}(153,447)
+ 2E_g(90,250)$, where the numbers in parentheses are phonon frequencies (in
cm$^{-1}$) determined for La$_2$NiO$_4$ by neutron scattering \cite{pint}.
The $A_{1g}$ lines are allowed in the $x^{\prime}x^{\prime}$ geometry, and
none are allowed in $x^{\prime}y^{\prime}$. The native IR-active phonon
lines are represented as $3A_{2u}(346,436,490)+4E_u(150,220,347,654)$.
Experimentally, the $A_{1g}$ and $E_u$ modes have been observed in previous
RLS \cite{burns,sug91} and FIR \cite{rice90,taj91} studies, respectively, on
undoped La$_2$NiO$_4$, and the frequencies obtained are in good agreement
with the neutron results \cite{pint}. Features at similar frequencies are
also prominent in measurements on our $x=0.225$ sample, as can be seen in
Figs.~2 and 3.

\begin{figure}
\centerline{\psfig{figure=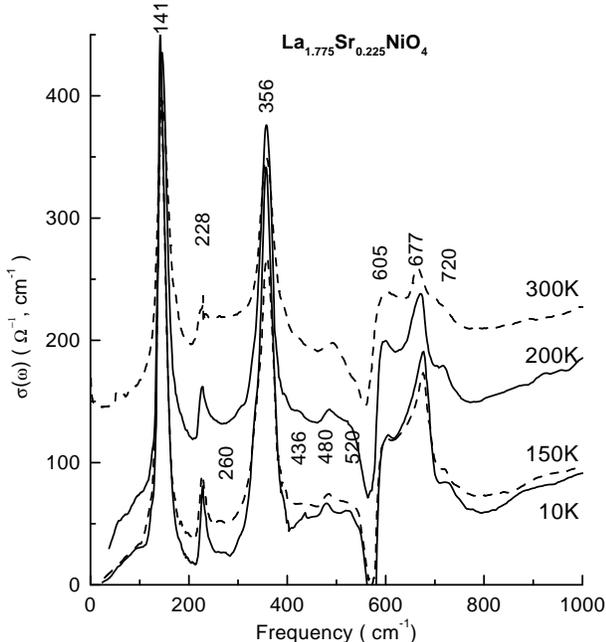,width=8cm,bbllx=14pt,bblly=245pt,%
bburx=550pt,bbury=825pt}} 
\caption{Temperature dependence of the
optical conductivity of La$_{1.775}$Sr$_{0.225}$NiO$_{4}$ in the
phonon spectral range.}
\end{figure}

The charge-stripe order is characterized (approximately \cite{1jt96,1jt98})
by the wave vector ${\bf Q}_c=(\epsilon,\epsilon,1)$ with $\epsilon=0.275$,
which has rotational symmetry {\bf mm2} with the second order axis along the
$x^{\prime}$-direction. The phonon states of the average structure at 
${\bf Q}_c$ are distributed among the IREPs of the {\bf mm2} wave vector
group in the following manner: 
$7\Sigma_1(x^{\prime})+3\Sigma_2+5\Sigma_3(y^{\prime})+6\Sigma_4(z)$. 
All of these states must be Raman-active, and $\Sigma_1$, $\Sigma_3$, and 
$\Sigma_4$ states are IR-active. The $\Sigma_1$ and $\Sigma_3$ modes must
appear in $x^{\prime}x^{\prime}$ and $x^{\prime}y^{\prime}$ RLS spectra,
respectively, if charge ordering creates lattice distortions belonging to the
$\Sigma_1$ IREP. Intensities of these new lines depend strongly on the type
and symmetry of the modes, and on the nature of the electron-phonon coupling.
Stripe order perturbations of the dynamical phonon matrix must also lead to a
lowering of the rotational symmetry of the phonon states at ${\bf Q}=0$ from
the $D_{4h}$ group to the $C_{2v}$ group, and to a splitting of the
two-fold degenerate $E$-states.

To identify the new peaks induced by stripe order, we make use of the
phonon-dispersion curves determined in the neutron-scattering study of 
La$_2$NiO$_4$ by Pintschovius {\it et al.} \cite{pint}. The $Q=0$
optically-active modes are expected to shift by only small amounts due to
doping, although some hardening is expected on cooling to low temperatures,
as the neutron study was performed at room temperature. Ideally, we would
like to compare with the dispersion curves along ${\bf Q}=(\xi,\xi,1)$;
however, since measurements have not been made along that direction, we will
compare with $(\xi,\xi,0)$, and rely on the fact that dispersion along
[001] is generally much smaller than in-plane dispersion \cite{pint}.

\begin{figure}
\centerline{\psfig{figure=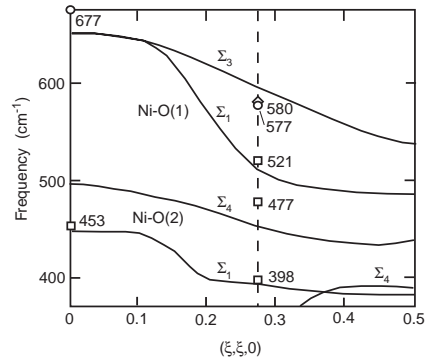,width=2.2in}}
\medskip
\caption{Comparison of major RLS and IR features in
La$_{1.775}$Sr$_{0.225}$NiO$_{4}$ at 5--10~K with the phonon
dispersions for bond-stretching modes measured for La$_2$NiO$_4$
at room temperature \protect\cite{pint}.  Squares: Raman $x'x'$;
diamond: Raman $x'y'$; circles: IR.  Dashed line indicates
charge-order wave vector, ${\bf Q}_c'$.}
\end{figure}

In Fig.~4, we show the highest-energy phonon branches along $(\xi,\xi,0)$
from Ref.~\cite{pint}. The two upper branches involve Ni-O(1) (in-plane)
bond-stretching motions, while the lower two derive from Ni-O(2) ($c$-axis)
bond-stretching. The dashed line indicates the in-plane charge-ordering wave
vector ${\bf Q}_c^{\prime}=(\epsilon,\epsilon,0)$. In the following, we will
associate the new features in the RLS and IR spectra induced by charge order
with ${\bf Q}_c^{\prime}$.

Starting with the Raman $x^{\prime}x^{\prime}$ spectra (Fig.~2), we observe
that, at 150~K and below, several peaks appear in the vicinity of
the native Ni-O(2) bond-stretching mode, which occurs at 447~cm$^{-1}$ for
$x=0$. A somewhat similar set of split peaks was observed for $x=\frac13$ in
the charge-ordered phase \cite{blum,yama98}. For $x=0.225$, significant peaks
appear at 398, 453, 477, and 521~cm$^{-1}$. We attribute the 453~cm$^{-1}$
feature to the $Q=0$ mode, and the 398~cm$^{-1}$ peak to the 
${\bf Q}_c^{\prime}$ mode, on the $\Sigma_1$ branch involving the already-noted
Ni-O(2) motion. The 477~cm$^{-1}$ peak appears to be close to the $\Sigma_4$
branch, which would require electron-phonon coupling to make it Raman
active. In contrast, the 521~cm$^{-1}$ peak is too high in energy to involve
Ni-O(2) motion (contrary to the speculation in Ref.~\cite{blum}). It must
correspond to the ${\bf Q}_c^{\prime}$ mode of the highest $\Sigma_1$
branch, which involves Ni-O(1) bond-stretching motion.

For the Raman $x^{\prime}y^{\prime}$ spectra, there is a
temperature-dependent peak at 580~cm$^{-1}$. Symmetry arguments, together
with the frequency, indicate that this is the ${\bf Q}_c^{\prime}$ mode of
the highest $\Sigma_3$ branch, which also involves Ni-O(1) bond-stretching
motion.

Finally, we arrive at the optical conductivity (Fig.~3). The peak at 
356~cm$^{-1}$ corresponds to the native $E_u$ mode involving Ni-O(1)-Ni
bond-bending motion. It was observed to split into at least 3 peaks in the
charge-ordered phase of $x=\frac13$ \cite{kats96}. The total splitting is
comparable to the width of our peaks. Model calculations that we have
performed using a modified rigid-ion model indicate that the splitting and
shifts of this peaks should be sensitive to the way in which the stripes are
positioned with respect to the lattice. The lack of a clear splitting in the
present case may indicate the absence of a unique positioning of the
stripes, which would be consistent with the incommensurate wave vector and
the finite correlation length for stripe order. Our calculations suggest
that the peak at 677~cm$^{-1}$ (corresponding to the 654~cm$^{-1}$ peak at 
$x=0$) should also be sensitive to stripe pinning, but, again, no clear
splitting is observed.

The most unusual feature in Fig.~3 is the strong dip at 577~cm$^{-1}$.
Because of the unusual nature of this apparent Fano antiresonance, the
measurement of the reflectivity in this frequency range was repeated several
times with different detectors and scanning rates in order to verify that it
is not an artifact. The occurrence of a dip instead of a peak clearly
indicates interference of a phonon mode with underlying electronic
conductivity. The energy of this feature uniquely associates it with Ni-O(1)
stretching motion.

Model calculations for stripes in a NiO$_2$ plane using the inhomogeneous
Hartree-Fock plus random-phase-approximation approach by Yi {\it et al.}
\cite{yi98} have demonstrated how the modulation of electronic hybridization
by Ni-O(1) bond-stretching modes results in extra phonon branches and strong
new IR modes. Our antiresonance mode is undoubtedly of this type. The model
calculations were done for stripes with a hole density of one per Ni site,
resulting in a large charge excitation gap \cite{yi98}. In contrast, for our
$x=0.225$ sample the hole density per Ni site along a stripe is 
$x/\epsilon=0.225/0.275=0.82$, which is consistent with a quasi-metallic
character. The antiresonant behavior demonstrates the existence of a finite
conductivity within the charge stripes at energies well below the pseudogap
(0.105 eV), which is presumably associated with charge motion transverse to
the stripes. The continued existence of the antiresonance at room
temperature, well above $T_c$, suggests that stripe correlations do not
disappear at $T_c$. Such a result should not be too surprising given that
the low-frequency conductivity remains small well above $T_c$, indicating
the continuing importance of strong correlations. The occurrence of stripe
correlations without static order has significant implications for
understanding the cuprates.

This work was supported in part by INTAS grant No. 96-0410 and
DFG/SFB 341 and SFFR research grant No.\ 2.4/247 of Ukraine. JMT
is supported by U.S. DOE Contract No.\ DE-AC02-98CH10886.


\begin{references}
\bibitem{ido91}  T. Ido, K. Magoshi, H. Eisaki, and S. Uchida, Phys. Rev. B
{\bf 44}, 12094 (1991).

\bibitem{kats96}  T. Katsufuji {\it et al.}, Phys. Rev. B {\bf 54}, R14230
(1996).

\bibitem{chen93}  C. H. Chen, S.-W. Cheong, and A. S. Cooper, Phys. Rev.
Lett. {\bf 71}, 2461 (1993).

\bibitem{tran94}  J. M. Tranquada, D. J. Buttrey, V. Sachan, and J. E.
Lorenzo, Phys. Rev. Lett. {\bf 73}, 1003 (1994).

\bibitem{lee97}  S.-H. Lee and S.-W. Cheong, Phys. Rev. Lett. {\bf 79}, 2514
(1997).

\bibitem{taj98}  S.Tajima {\it et al.}, J. Phys. Chem. Solids {\bf 59}, 2015
(1998).

\bibitem{1jt95}  J. M. Tranquada {\it et al.}, Nature {\bf 375}, 561 (1995).

\bibitem{zim98}  M. v. Zimmermann {\it et al.}, Europhys. Lett. {\bf 41},
629 (1998).

\bibitem{tran96}  J. M. Tranquada {\it et al.}, Phys. Rev. B {\bf 54}, 7489
(1996).

\bibitem{1jt96}  J. M. Tranquada, D. J. Buttrey, and V. Sachan, Phys.Rev.B
{\bf 54}, 12318 (1996).

\bibitem{vigl97}  A. Vigliante {\it et al.}, Phys. Rev. B {\bf 56}, 8248
(1997).

\bibitem{2jt97}  J. M. Tranquada {\it et al.}, Phys.Rev.Lett. {\bf 78}, 338
(1997).

\bibitem{ost97}  J. E. Ostenson {\it et al.}, Phys.Rev. B {\bf 56}, 2820
(1997).

\bibitem{1zaa99}  H. Eskes {\it et al.}, Phys. Rev. B {\bf 58}, 6963 (1998).

\bibitem{kive98}  S. A. Kivelson, E. Fradkin, and V. J. Emery, Nature 
{\bf 393}, 550 (1998).

\bibitem{rami96}  A. P. Ramirez {\it et al.}, Phys. Rev. Lett. {\bf 76}, 447
(1996).

\bibitem{mcqu99}  R. J. McQueeney, J. L. Sarrao, and R. Osborn, Phys. Rev. B
{\bf 60}, 80 (1999).

\bibitem{blum}  G. Blumberg, M. V. Klein, and S. W. Cheong, Phys. Rev. Lett.
{\bf 80}, 564, (1998).

\bibitem{yama98}  K. Yamamoto, T. Katsufuji, T. Tanabe, and Y. Tokura, Phys.
Rev. Lett. {\bf 80}, 1493 (1998).

\bibitem{pint}  L.Pintschovius {\it et al.}, Phys. Rev. B {\bf 40}, 2229
(1989).

\bibitem{burns}  G. Burns {\it et al.}, Phys. Rev. B {\bf 42}, 19777 (1990).

\bibitem{sug91}  S. Sugai {\it et al.}, Physica C {\bf 185-189}, 895 (1991).

\bibitem{rice90}  D. E. Rice, M. K. Crawford, D. J. Buttrey, and W. E.
Farneth, Phys. Rev. B {\bf 42}, 8787 (1990).

\bibitem{taj91}  S.Tajima {\it et al.}, Phys. Rev. B {\bf 43}, 10496 (1991).

\bibitem{1jt98}  P. Wochner, J. M. Tranquada, D. J. Buttrey, and V. Sachan,
Phys. Rev. B {\bf 57}, 1066 (1998).

\bibitem{yi98}  Y.-S. Yi, Z.-G. Yu, A. R. Bishop, and J. T. Gammel, Phys.
Rev. B {\bf 58}, 503 (1998).
\end{references}
\end{document}